\documentclass[10pt,aps,prb,showpacs]{revtex4}
\usepackage{eurosym}
\usepackage{amsmath,amssymb}
\usepackage{amsfonts}
\usepackage{color}
\usepackage{epsfig}
\usepackage{graphicx}
\usepackage{epstopdf}

\begin{document}

\title{Interfaces between Bose-Einstein and Tonks-Girardeau atomic gases}
\author{Giovanni Filatrella$^{1}$ }
\author{Boris A. Malomed$^{2}$ }
\affiliation{$^1$ Department of Sciences and Technologies of the University of Sannio and
CNISM unit Salerno, I-82100 Benevento, Italy}
\affiliation{$^2$ Department of Physical Electronics, School of Electrical Engineering,
Faculty of Engineering, Tel Aviv University, Tel Aviv 69978, Israel}
\date{\today }

\begin{abstract}
We consider one-dimensional mixtures of an atomic Bose-Einstein condensate
(BEC) and Tonks-Giradeau (TG) gas. The mixture is modeled by a coupled
system of the Gross-Pitaevskii equation for the BEC and the quintic
nonlinear Schr\"{o}dinger equation for the TG component. An immiscibility
condition for the binary system is derived in a general form. Under this
condition, three types of BEC-TG interfaces are considered: domain walls
(DWs) separating the two components; bubble-drops (BDs), in the form of a
drop of one component immersed into the other (BDs may be considered as
bound states of two DWs); and bound states of bright and dark solitons
(BDSs). The same model applies to the copropagation of two optical waves in
a colloidal medium. The results are obtained by means of systematic
numerical analysis, in combination with analytical Thomas-Fermi
approximations (TFAs). Using both methods, families of DW states are
produced in a generic form. BD complexes exist solely in the form of a TG
drop embedded into the BEC background. On the contrary, BDSs exist as bound
states of TG bright and BEC dark components, and vice versa.
\end{abstract}

\pacs{03.75.Nt, 03.75.Mn, 05.30.Jp}
\maketitle

\section{ Introduction}

Binary systems, whose behavior crucially depends on the underlying condition
of immiscibility or miscibility \cite{Mineev}, play a fundamentally
important role in many areas of physics. In the case of immiscibility, a
major effect is the formation of domain walls (DWs) between regions occupied
by immiscible components. Commonly known are DWs in media featuring a
vectorial order parameter, such as ferromagnets \cite{magnetic},
ferroelectrics \cite{electric}, and liquid crystals \cite{liquid}. In
self-defocusing optical media, DWs separate regions occupied by
electromagnetic waves with orthogonal circular polarizations of light \cite%
{Haelt,me}. Similar interface patterns were predicted in arrays of nonlinear
optical waveguides, modeled by discrete nonlinear Schr\"{o}dinger equations
(NLSEs) \cite{Panos}.

DWs are known in superfluids too, where they are formed by immiscible binary
Bose-Einstein condensates (BECs), as predicted theoretically \cite{Warsaw}
and demonstrated in experiments \cite{BEC-experiment}. In the mean-field
approximation \cite{book}, such settings are modeled by systems of
nonlinearly coupled Gross-Pitaevskii equations (GPEs) with the cubic
self-repulsive nonlinearity, which are similar to coupled NLSEs describing
the above-mentioned optical DWs \cite{Haelt,me}. In their stationary form,
these equations coincide with coupled cubic Ginzburg-Landau equations
modeling DWs in dissipative patterns, such as interfaces between rolls with
different orientations in large-area Rayleigh-Benard convection \cite{MNT}.

The analysis of the DWs in BEC was extended for broader settings, including
linear interconversion between the immiscible components\ (this is possible
when they represent two different hyperfine states of the same atom coupled
by a resonant radiofrequency wave) \cite{BEClin-coupling}, dipolar \cite{DD}
and spinor (three-component) condensates \cite{spinor}, as well as the BEC
discretized by trapping in a deep optical-lattice (OL) potentials \cite%
{Mering}. Furthermore, the study of the DWs was recently extended for
immiscible binary BECs\ with three-particle collisions \cite{we}, in the
case when the related losses may be neglected, the respective coupled GPEs
featuring the cubic-quintic repulsive nonlinearity \cite{CQ}.

In the effectively one-dimensional (1D) setting, ultracold bosonic gases
with strong inter-atomic repulsion may be cast in the Tonks-Girardeau (TG)
state, which emulates the gas of non-interacting fermions \cite{TG},
provided that the energy of the repulsive interaction between bosons exceeds
their kinetic energy, while the opposite situation corresponds to the BEC\
phase in the bosonic gas (a review of the TG model was given in Ref. \cite{3}%
). The TG gas of hard-core bosons has been realized experimentally, using
tight transverse confinement \cite{1,1b}. In particular, a longitudinal OL
potential was used to increase the effective mass in the trapped state, thus
making the kinetic energy small enough \cite{1}.

It is commonly known that GPEs furnish very accurate description of the BEC
in atomic gases. A similar macroscopic model of the TG gas is offered by the
NLSE with the quintic self-repulsion term \cite{Kolo}. In a rigorous form,
the relevance of the corresponding sextic term in the free-energy density of
the three-dimensional bosonic gas in its ground state, which reduces to the
quasi-1D TG phase, was demonstrated in Ref. \cite{Lieb}, under condition $%
\mathcal{G}\mathcal{L}\gg \mathcal{N}$, where $\mathcal{G}$, $\mathcal{L}$,
and $\mathcal{N}$ are, respectively, the inter-atomic repulsion strength,
system's length, and the total number o atoms. The quintic model was used in
various contexts, including shock waves \cite{shock}, dark \cite{dark} and
gap-mode \cite{gap-sol} solitons, as well as bright solitons supported by
dipole-dipole interactions \cite{dip-dip}, and, recently, DWs in immiscible
binary TG gases \cite{we}. Further, oscillation frequencies derived from
fermionic hydrodynamic equations, which apply to the hard-core TG gas, were
found to be close to their counterparts predicted by the quintic NLSE \cite%
{hydro}. Coupled quintic NLSEs also arise in works aimed at constructing the
ground state of a binary TG mixture in the harmonic-oscillator potential by
means of the density-functional method \cite{density-functional}. On the
other hand, this approach may not apply to TG gases beyond the framework of
static configurations and hydrodynamic regimes. In particular, it fails for
strongly non-equilibrium problems, such as merger of distinct gas clouds
\cite{Girardeau}.

As concerns the mixtures, it may be interesting to consider binary systems
including the TG gas and another quantum-gas component. In particular, exact
solutions\ were found for the ground state of TG-Fermi mixtures \cite%
{Minguzzi}. The binary gas of impenetrable bosons is solvable too \cite%
{cTG-BA}. The objective of the present work is to introduce basic nonlinear
complexes, such as DWs, bubble-drop (BD) modes (bound pairs of two DWs), and
dark-bright solitons (DBSs), in an immiscible system of TG and BEC gases. In
the experiment, the system may be realized, in particular, as a bosonic gas
composed of two atomic species under tight transverse confinement, with a
longitudinal OL potential acing on (being relatively close to a resonance
with) one species only. Then, as the experimental setting presented in Ref.
\cite{1} suggests, a large effective mass of the near-resonant component
will bring it into the TG state, while the other component may stay in the
BEC phase.

As a model for this system, in Section II we adopt the cubic GPE for the
self-repulsive BEC component coupled by the cubic (collisional) repulsive
term to the quintic NLSE for the TG species. The use of the latter equation
is appropriate, as we study only static configurations of the system. The
same model may find a realization in optics as a model of colloidal
waveguides. It has been recently demonstrated that, selecting the size of
metallic nanoparticles in the colloid and their concentration, one can
engineer desirable coefficients of the corresponding cubic and quintic
nonlinearity \cite{Cid}. In particular, it is possible to design a waveguide
which features a nearly pure quintic nonlinearity at a particular
wavelength, while the cubic response dominates at a different wavelength.
The copropagation of optical signals carried by these wavelengths will then
emulate the TG-BEC system.

DWs separating the BEC and TG phases are addressed in Section III. We derive
the respective immiscibility condition (see Eq. (\ref{g<}) below), and then
generate DW states in a systematic way, using both numerical solutions and
the analytical Thomas-Fermi approximation (TFA). The latter method makes it
possible to obtain some DWs in an explicit analytical form, as given below
by Eqs. (\ref{exact}) - (\ref{WTFA}). DB and DBS complexes are considered in
Sections IV and V, respectively, again using a combination of analytical and
numerical methods. The analysis predicts that the DB states exist solely in
the form of the TG drop embedded into the BEC background (bubble), but not
in the opposite case; on the other hand, the DBS are predicted in either
case of the bright TG soliton embedded into the BEC dark soliton, or vice
versa. These predictions are fully corroborated by numerical results. The
paper is concluded by Section VI.

\section{The model}

The system of coupled NLSEs with the cubic and quintic nonlinear terms for
the BEC and TG components, $\psi _{1}$ and $\psi _{2}$, with respective
scaled masses $m_{1}$ and
\begin{equation}
m_{2}\equiv mm_{1}  \label{mmm}
\end{equation}%
(i.e., $m$ is the relative effective mass of TG components which, as said
above, may be made larger than the actual atomic mass \cite{1}) is
\begin{eqnarray}
&&i\frac{\partial \Psi _{1}}{\partial T}=-\frac{1}{2m_{1}}\frac{\partial
^{2}\Psi _{1}}{\partial X^{2}}+\left( \gamma |\Psi _{1}|^{2}+\Gamma |\Psi
_{2}|^{2}\right) \Psi _{1},  \notag \\
&&  \label{first} \\
&&i\frac{\partial \psi _{2}}{\partial T}=-\frac{1}{2mm_{1}}\frac{\partial
^{2}\Psi _{2}}{\partial X^{2}}+\left( \left\vert \Psi _{2}\right\vert
^{4}+\Gamma |\Psi _{1}|^{2}\right) \Psi _{2},\;\;\;  \notag
\end{eqnarray}%
where real parameters $\gamma \geq 0$ and $\Gamma >0$ are strengths of the
self-repulsion of the BEC component, and repulsion between the BEC and TG
ones, respectively, while $\hbar $ and the coefficient of the effective
quintic self-repulsion of the TG component are scaled to be $1$ (in the
notation of Ref. \cite{Kolo}, the natural value of the latter coefficient is
$\pi ^{2}$). Substitution
\begin{equation}
\Psi _{1,2}\equiv \sqrt{\Gamma }\psi _{1,2},~T\equiv t/\Gamma ^{2},~X\equiv
x/\left( \sqrt{m_{1}}\Gamma \right) ,\gamma \equiv \Gamma g  \label{rescale}
\end{equation}%
makes it possible to further fix $\Gamma \equiv 1$ and $m_{1}\equiv 1$, thus
simplifying Eq. (\ref{first}) to a system with two free coefficients, $m$
and $g$:%
\begin{eqnarray}
&&i\frac{\partial \psi _{1}}{\partial t}=-\frac{1}{2}\frac{\partial ^{2}\psi
_{1}}{\partial x^{2}}+\left( g|\psi _{1}|^{2}+|\psi _{2}|^{2}\right) \psi
_{1},  \label{BEC} \\
&&i\frac{\partial \psi _{2}}{\partial t}=-\frac{1}{2m}\frac{\partial
^{2}\psi _{2}}{\partial x^{2}}+\left( \left\vert \psi _{2}\right\vert
^{4}+|\psi _{1}|^{2}\right) \psi _{2}.\;\;\;  \label{TG}
\end{eqnarray}

For the above-mentioned spatial-domain optical model (with $t$ replaced by
the propagation distance, $Z$, and transverse coordinate $X$), the scaled
propagation equations are derived, using the standard procedure \cite{KA}, as%
\begin{eqnarray}
&&i\frac{\partial \Psi _{1}}{\partial Z}=-\frac{1}{2}\frac{\partial ^{2}\Psi
_{1}}{\partial X^{2}}+\left( G_{3}|\Psi _{1}|^{2}+|\Psi _{2}|^{2}\right)
\Psi _{1},  \notag \\
&&  \label{omega} \\
&&i\frac{\partial \Psi _{2}}{\partial Z}=-\frac{1}{2k}\frac{\partial
_{2}^{2}\Psi }{\partial X^{2}}+\frac{\omega ^{2}}{k}\left( G_{5}\left\vert
\Psi _{2}\right\vert ^{4}+|\Psi _{1}|^{2}\right) \Psi _{2}.  \notag
\end{eqnarray}%
Here $\Psi _{1}$ and $\Psi _{2}$ represent amplitudes of the copropagating
electromagnetic waves with relative wavenumber and frequency $k=k_{2}/k_{1}$
and $\omega =\omega _{2}/\omega _{1}$, cf. Eq. (\ref{mmm}). Further, $%
G_{3,5} $ in Eq. (\ref{omega}) are the cubic and quintic SPM coefficients
for the two waves, while the cubic XPM coefficient is normalized to be $1$.
Additional rescaling,%
\begin{gather}
\Psi _{1}\equiv \frac{k}{\omega ^{2}}\sqrt{\frac{1}{G_{5}}}\psi _{1},~\Psi
_{2}\equiv \frac{1}{\omega }\sqrt{\frac{k}{G_{5}}}\psi _{2},  \notag \\
Z\equiv \frac{\omega ^{2}G_{5}}{k}t,~X\equiv \omega \sqrt{\frac{G_{5}}{k}}%
x,~G_{3}\equiv \frac{\omega ^{2}}{k}g,  \notag
\end{gather}%
transforms Eq. (\ref{omega}) into the system of equations (\ref{BEC}) and (%
\ref{TG}), with $m\equiv k$.

The Hamiltonian corresponding to Eqs. (\ref{BEC}) and (\ref{TG}) is
\begin{equation}
H=\int_{-\infty }^{+\infty }\left( \frac{1}{2}\left\vert \frac{\partial \psi
_{1}}{\partial x}\right\vert ^{2}+\frac{1}{2m}\left\vert \frac{\partial \psi
_{2}}{\partial x}\right\vert ^{2}+\frac{g}{2}|\psi _{1}|^{4}+\frac{1}{3}%
|\psi _{2}|^{6}+\left\vert \psi _{1}\right\vert ^{2}\left\vert \psi
_{2}\right\vert ^{2}\right) dx.  \label{H}
\end{equation}%
In addition to $H$, the system preserves the norms (scaled numbers of atoms
in the ultracold gas, or total powers of the two waves, in terms of the
optical model),
\begin{equation}
N_{1,2}=\int_{-\infty }^{+\infty }|\psi _{1,2}(x)|^{2}dx,  \label{N}
\end{equation}%
of the two components, and, for dynamical solutions, also the total
momentum, $P=i\int_{-\infty }^{+\infty }\sum_{n=1}^{2}\psi _{n}\left(
\partial \psi _{n}^{\ast }/\partial x\right) dx$, although, as mentioned
above, the use of the quintic NLS equation for the description of dynamics
of the TG gas may be impugnable.

Stationary solutions to Eqs. (\ref{BEC}) and (\ref{TG}) with positive
chemical potentials $\mu _{1,2}$ are looked for as
\begin{equation}
\psi _{1,2}(x,t)\equiv \exp \left( -i\mu _{1,2}t\right) \phi _{1,2}(x),
\label{mu}
\end{equation}%
with real functions $\phi _{1,2}(x)$ satisfying equations
\begin{gather}
\mu _{1}\phi _{1}=-\frac{1}{2}\frac{d^{2}\phi _{1}}{dx^{2}}+\left( g\phi
_{1}^{2}+\phi _{2}^{2}\right) \phi _{1},  \label{phi1} \\
\mu _{2}\phi _{2}=-\frac{1}{2m}\frac{d^{2}\phi _{2}}{dx^{2}}+\left( \phi
_{2}^{4}+\phi _{1}^{2}\right) \phi _{2}\;\;  \label{phi2}
\end{gather}%
(in the optical model, $-\mu _{1,2}$ represent the propagation constants of
the two waves). Equations (\ref{phi1}) and (\ref{phi2}), if considered as
equations of the evolution along $x$, conserve the formal Hamiltonian,%
\begin{equation}
h=\frac{1}{2}\left( \frac{d\phi _{1}}{dx}\right) ^{2}+\frac{1}{2m}\left(
\frac{d\phi _{2}}{dx}\right) ^{2}+\mu _{1}\phi _{1}^{2}+\mu _{2}\phi
_{2}^{2}-\frac{g}{2}\phi _{1}^{4}-\frac{1}{3}\phi _{2}^{6}-\phi _{1}^{2}\phi
_{2}^{2},  \label{h}
\end{equation}%
cf. Eq. (\ref{H}).

Normalizing the stationary wave functions as%
\begin{equation}
\phi _{1}\equiv \mu _{2}^{1/2}\tilde{\phi}_{1},~\phi _{2}\equiv \mu
_{2}^{1/4}\tilde{\phi}_{2},~x\equiv \mu _{2}^{-1/4}\tilde{x},~m\equiv \mu
_{2}^{-1/2}\tilde{m},~\mu _{1}\equiv \mu _{2}^{-1/2}\tilde{\mu}_{1},
\label{tilde}
\end{equation}%
one may fix $\mu _{2}=1$, which implies that the density of the uniform TG
component is also fixed to be $1$, see Eq. (\ref{bc}) below. Numerical
results for DWs are presented in the following section chiefly for this
case, while in Sections IV and V\ other normalizations are used for BD and
DBS states.

\section{Domain walls}

\subsection{Analytical considerations}

Solutions of stationary equations (\ref{phi1}) and (\ref{phi2}) for the
domain wall (DW) separating semi-infinite domains occupied by the BEC and TG
components (or the spatial domains occupied by the copropagating waves in
the above-mentioned optics model) are specified by the following boundary
conditions (b.c.), which include the respective asymptotic densities, $\phi
_{1}^{2}\left( x=-\infty \right) \equiv \left( n_{1}\right) _{\mathrm{asympt}%
}$ and $\phi _{2}^{2}\left( x=+\infty \right) \equiv \left( n_{2}\right) _{%
\mathrm{asympt}}$:
\begin{eqnarray}
\left( n_{1}\right) _{\mathrm{asympt}} &=&\mu _{1}/g,~\phi _{2}\left(
x=-\infty \right) =0,  \notag \\
&&  \label{bc} \\
\phi _{1}\left( x=+\infty \right) &=&0,~\left( n_{2}\right) _{\mathrm{asympt}%
}=\sqrt{\mu _{2}}.  \notag
\end{eqnarray}%
The condition that formal Hamiltonian (\ref{h}) must take the same values at
$x\rightarrow -\infty $ and $x\rightarrow +\infty $ across the solution
imposes a restriction on b.c. (\ref{bc}), which relates the two chemical
potentials:%
\begin{equation}
\mu _{1}^{2}=\left( 4g/3\right) \mu _{2}^{3/2}.  \label{mumu}
\end{equation}%
In terms of the asymptotic densities, condition (\ref{bc}) takes the form of%
\begin{equation}
\left( n_{1}\right) _{\mathrm{asympt}}=\left( 2/\sqrt{3g}\right) \left(
n_{2}\right) _{\mathrm{asympt}}^{3/2}~,  \label{nn}
\end{equation}%
which actually implies the balance of the pressure applied to the DW from
the two sides. Naturally, the mass ratio ($m$) does not appear in Eqs. (\ref%
{mumu}) and (\ref{nn}). Note that, if condition (\ref{nn}) between the
densities does not hold initially, the DW in a finite system, which
corresponds to real experimental settings, will move to a position at which
the condition holds for the accordingly modified densities.

Further, the \emph{immiscibility condition} for the BEC\ and TG, which is
necessary for the existence of the DW separating the two quantum gases, is
that the demixed configuration must provide a \emph{smaller} energy density
(defined as per Eq. (\ref{H})) than a uniformly mixed state with densities
which are equal to half of asymptotic densities (\ref{bc}):%
\begin{equation}
\left( n_{1}\right) _{\mathrm{uni}}=\frac{1}{2}\left( n_{1}\right) _{\mathrm{%
asympt}}\equiv \frac{\mu _{1}}{2g},~\left( n_{2}\right) _{\mathrm{uni}}=%
\frac{1}{2}\left( n_{2}\right) _{\mathrm{asympt}}\equiv \frac{\sqrt{\mu _{2}}%
}{2}.  \label{uni}
\end{equation}%
The uniform state corresponds to values of the chemical potentials different
from $\mu _{1}$ and $\mu _{2}$, namely, $\left( \mu _{1}\right) _{\mathrm{uni%
}}=g\left( n_{1}\right) _{\mathrm{uni}}+\left( n_{2}\right) _{\mathrm{uni}}$%
, $\left( \mu _{2}\right) _{\mathrm{uni}}=\left( n_{2}\right) _{\mathrm{uni}%
}^{2}+\left( n_{2}\right) _{\mathrm{uni}}$. Then, the comparison of the
average energy densities of the demixed and uniform states gives rise to the
immiscibility condition in the following form:
\begin{equation}
\frac{\mu _{1}^{2}}{g}+\mu _{2}^{3/2}<\frac{2\mu _{1}\sqrt{\mu _{2}}}{g}.
\label{<}
\end{equation}%
Further, the substitution of relation (\ref{mumu}) in Eq. (\ref{<})
transforms it into an inequality for the self-repulsion strength of the BEC
component, $g$:
\begin{equation}
g<\frac{48}{49}\frac{1}{\sqrt{\mu _{2}}}=\frac{48}{49}\frac{1}{\left(
n_{2}\right) _{\mathrm{asympt}}}\equiv g_{\max }.  \label{g<}
\end{equation}%
In particular, for $\mu _{2}=1$ (as said above, this value will be fixed by
rescaling), Eq. (\ref{g<}) yields $g_{\max }=48/49\approx \allowbreak 0.9796$%
.

The DW may be characterized by an effective width of its core, which may be
naturally defined by the following integral expression:
\begin{equation}
W=\frac{1}{\left( n_{1}\right) _{\mathrm{asympt}}\left( n_{2}\right) _{%
\mathrm{asympt}}}\int_{-\infty }^{+\infty }\phi _{1}^{2}(x)\phi
_{2}^{2}(x)dx.  \label{W}
\end{equation}%
The analysis should produce $W$ as a function of parameters $m$ and $g$,
once the TG asymptotic density is fixed by setting $\mu _{2}=1$, see Fig. %
\ref{fig:Wvsm} below.

Close to the existence boundary of the DW, i.e., at $0<g_{\max }-g\ll
g_{\max }$, the DW becomes very wide. In this limit, the dependence between
the width and proximity to the threshold may be estimated as follows: the
density of the gradient energy in Eq. (\ref{H}) scales as $1/W^{2}$, hence
the full gradient energy scales as $1/W$, and the respective effective force
may be estimated as $-\partial H/\partial W$ $\sim 1/W^{2}$. It must be
balanced by the effective bulk force vanishing at $g=g_{\max }$, which
scales as $g_{\max }-g$. Thus, the equilibrium condition predicts that the
DW's width diverges at $g_{\max }-g\rightarrow 0$ as%
\begin{equation}
W\sim 1/\sqrt{g_{\max }-g}.  \label{broad}
\end{equation}

Accurate analytical results for the DW can be obtained in the limit case
corresponding to $g\gg 1$, or to $m\ll 1$, when the derivative term in Eq. (%
\ref{phi1}) may be neglected (which actually implies the use of the TFA \cite%
{book}), yielding%
\begin{equation}
\phi _{1}^{2}=\frac{\mu _{1}-\phi _{2}^{2}}{g}.  \label{1/g}
\end{equation}%
Strictly speaking, the assumption of $m\ll 1$ may contradict the
experimentally relevant way of the realization of the TG gas, based on
making the respective effective mass large \cite{1}. Nevertheless, the other
option justifying the applicability of the TFA, $g\gg 1$, is quite relevant.

The substitution of expressions (\ref{1/g}) and (\ref{mumu}) in Eq. (\ref%
{phi2}) leads to the single stationary equation for $\phi _{2}$, with the
cubic-quintic nonlinearity:%
\begin{equation}
\mu _{2}^{3/4}\left( \mu _{2}^{1/4}-\frac{2}{\sqrt{3g}}\right) \phi _{2}+%
\frac{1}{2m}\frac{d^{2}\phi _{2}}{dx^{2}}-\phi _{2}^{5}+\frac{1}{g}\phi
_{2}^{3}=0  \label{single}
\end{equation}%
(Eq. (\ref{mumu}) was used here to eliminate $\mu _{1}$). Then, using a
known particular exact solution for the DW solution of Eq. (\ref{single})
\cite{Zeev}, we obtain a solution at the following values of the chemical
potentials:%
\begin{equation}
\left( \mu _{2}\right) _{\mathrm{TFA}}=\left( \frac{3}{4g}\right)
^{2},~\left( \mu _{1}\right) _{\mathrm{TFA}}=\frac{3}{4g},  \label{exact}
\end{equation}%
i.e., $\left( n_{1}\right) _{\mathrm{asympt}}=3/\left( 4g^{2}\right) $, $%
\left( n_{2}\right) _{\mathrm{asympt}}=3/\left( 4g\right) $, in the form of%
\begin{equation}
\left( \phi _{2}(x)\right) _{\mathrm{TFA}}=\frac{1}{2}\sqrt{\frac{3}{g}\frac{%
1}{1+\exp \left( \pm \sqrt{3m/2}x/g\right) }}.  \label{DW}
\end{equation}%
Note that $\left( \mu _{2}\right) _{\mathrm{DW}}$ satisfies condition (\ref%
{g<}), and rescaling (\ref{tilde}) transforms values (\ref{exact}) into $%
\left( \mu _{2}\right) _{\mathrm{DW}}=\left( \mu _{1}\right) _{\mathrm{DW}}=1
$. The width of this solution, defined according to Eq. (\ref{W}), is%
\begin{equation}
W_{\mathrm{TFA}}=\sqrt{2/\left( 3m\right) }g.  \label{WTFA}
\end{equation}%
Comparison of analytical solution (\ref{DW}) with its numerical counterpart
is displayed below in Fig. \ref{fig:exact}.

An analytical approach can also be developed in the opposite limit of $%
m\rightarrow \infty $ (very heavy TG atoms), when the TFA may be applied to
Eq. (\ref{phi2}) (i.e., the derivative term may be dropped in this
equation), reducing it to%
\begin{equation}
\phi _{2}^{2}(x)=\sqrt{\mu _{2}-\phi _{1}^{2}(x)}.  \label{phi2phi1}
\end{equation}%
In this case, Eq. (\ref{phi1}) takes the form of%
\begin{equation}
\mu _{1}\phi _{1}=-\frac{1}{2}\frac{d^{2}\phi _{1}}{dx^{2}}+g\phi _{1}^{3}+%
\sqrt{\mu _{2}-\phi _{1}^{2}}\phi _{1},  \label{phi1nophi2}
\end{equation}%
cf. Eqs. (\ref{1/g}) and (\ref{single}) derived above in the opposite limit
of $m\rightarrow 0$. In the particular case of $g=0$ (no intrinsic
interaction in the BEC component), Eq. (\ref{phi1nophi2}) coincides with the
equation considered in Ref. \cite{Shabtay}, in the context of a completely
different physical model, for a resonantly absorbing Bragg reflector in
optics. The formal Hamiltonian for Eq. (\ref{phi1nophi2}) is
\begin{equation}
h=\frac{1}{2}\left( \frac{d\phi _{1}}{dx}\right) ^{2}+\mu _{1}\phi _{1}^{2}-%
\frac{g}{2}\phi _{1}^{4}-\frac{2}{3}\left( \mu _{2}-\phi _{1}^{2}\right)
^{3/2},  \label{phi1h}
\end{equation}%
cf. Eq. (\ref{h}). The DW solution corresponds to a solution of Eq. (\ref%
{phi1nophi2}) with the b.c. produced by Eq. (\ref{bc}): $\phi _{1}^{2}\left(
x=-\infty \right) =\mu _{1}/g,~\phi _{1}\left( x=+\infty \right) =0$, and,
simultaneously, $\phi _{1}^{2}\left( x=-\infty \right) =\mu _{2}$, the
latter relation following from Eq. (\ref{phi2phi1}. Obviously, this b.c. set
is self-consistent solely for $\mu _{1}=g\mu _{2}$. In the combination with
this relation, equating values of $h$ corresponding to $x=-\infty $ and $%
x=+\infty $, as per Eq. (\ref{phi1h}), yields $g\sqrt{\mu _{2}}=-4/3$, which
is impossible, as $g$ cannot be negative. Thus, the DW solution does not
exist in this approximation. Nevertheless, it readily produces analytical
solutions for complexes of other types, BD and DBS, as shown below.

\subsection{Numerical findings}

\begin{figure}[tbph]
\centerline{} \includegraphics[width=0.35\textwidth]{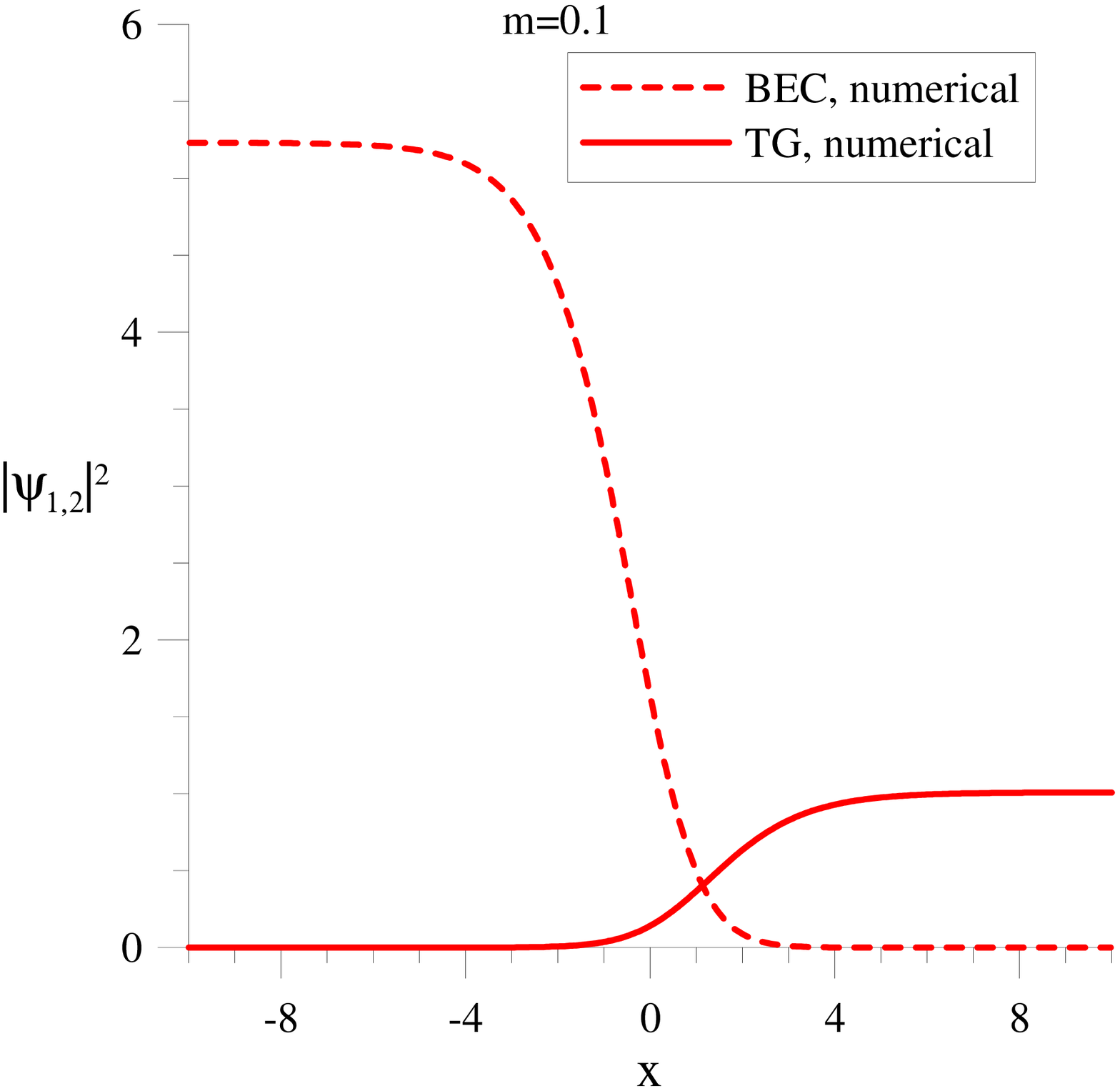}
\includegraphics[width=0.35
\textwidth]{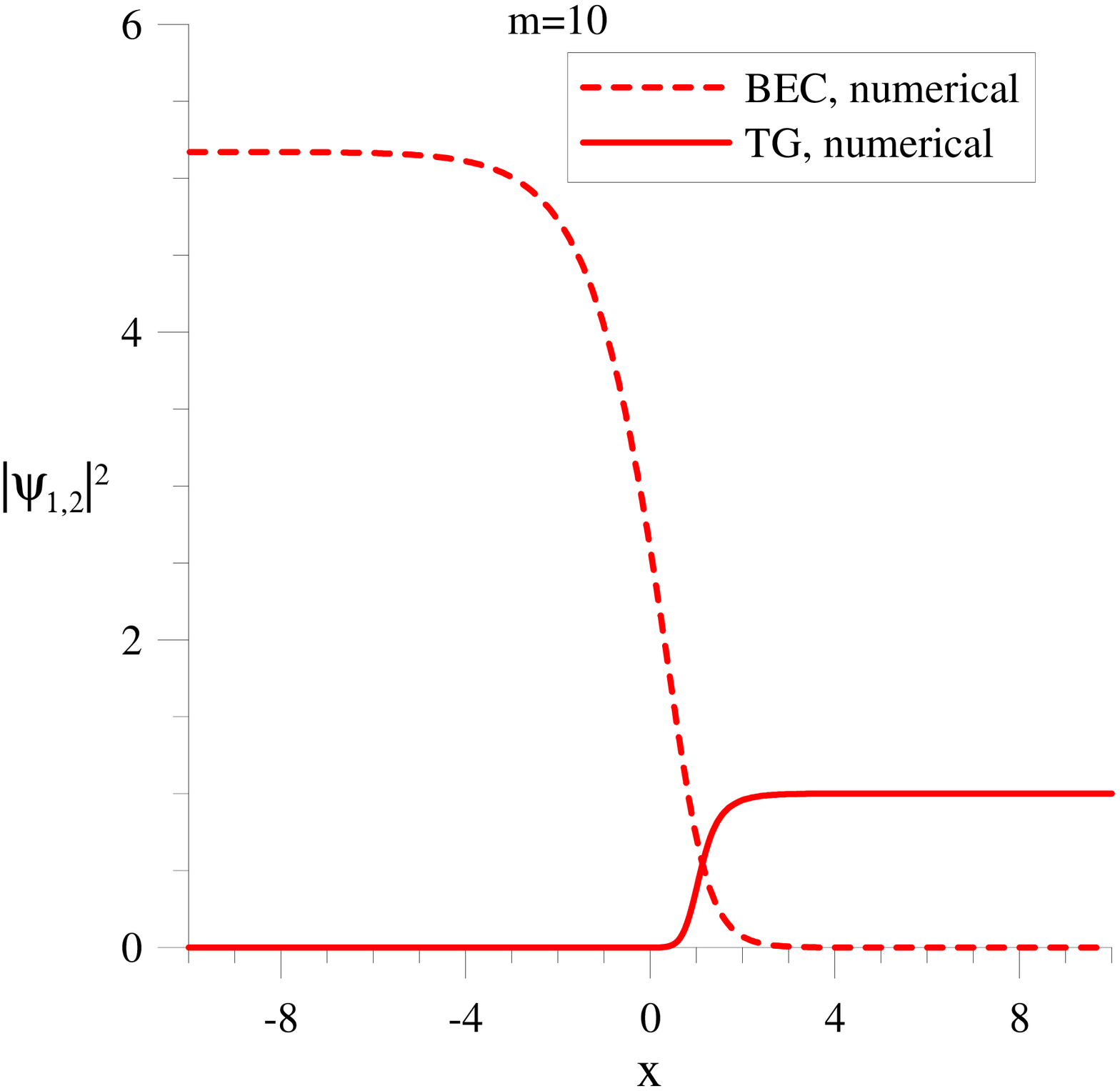} \centerline{(a) \hspace{6cm} (b)}
\caption{(Color online) DW density profiles produced by the imaginary-time
solution of Eqs. (\protect\ref{BEC}), (\protect\ref{TG}), and subsequently
verified by the real-time propagation, for $\protect\mu _{2}=1$, $\protect%
\mu _{1}=0.258$, $g=0.05$, and two values of the relative mass, $m=0.1$ and $%
10$. In terms of the optics model, the BEC and TG densities correspond to
power densities of the waves subject to the action of the cubic and quintic
self-defocusing nonlinearity, respectively (in captions to the following
figures, they are referred to as ``cubic" and ``quintic" components,
respectively).}
\label{fig:DW}
\end{figure}

Generic DW structures obtained from numerical solutions of Eqs. (\ref{BEC})
and (\ref{TG}) are displayed in Fig. \ref{fig:DW}. Stationary solutions have
been obtained by means of the imaginary-time-propagation method. The
validity and stability of the solutions was then checked by simulations in
real time. The simulations in both imaginary and real time were run by means
of the split-step Crank-Nicholson method \cite{Adhikari02}.

Stable solutions have been found in a wide range of values of $m$ at $g<1$,
and no solutions could be obtained at $g>1$, which is readily explained by
Eq. (\ref{g<}) (recall we set $\mu _{2}=1$). Actual numerical results for
the DWs have been obtained for $g\leq 0.85$. In other words, a natural
conclusion is that the DW exists if the BEC-TG repulsion is stronger than
the intrinsic repulsion of the BEC component. In the remaining interval of $%
0.85<g<48/49$ (see Eq. (\ref{g<})), it is difficult to generate DWs
numerically, as the initial guess for finding the solution by means of the
imaginary-time propagation should be very close to the true one, in case the
solution is sought for close to its existence boundary. Unlike $g$, the
dependence of the DW solutions on $m$ is very weak, as seen in Fig. \ref%
{fig:DW} which displays examples for $m=0.1$ and $m=10$ (the comparison of
the results obtained for large and small values of $m$, presented here and
below, is appropriate even if very small values of $m$ may be experimentally
irrelevant, as mentioned above).

\begin{figure}[tbph]
\includegraphics[width=0.45\textwidth]{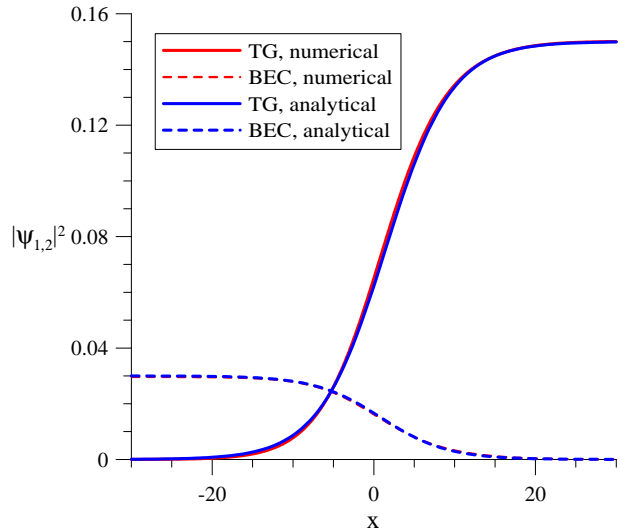}
\caption{(Color online) The DW density profile, produced by the numerical
solution of Eqs. (\protect\ref{BEC}) and (\protect\ref{TG}), and its
comparison with the analytical prediction (TFA), given by Eqs. (\protect\ref%
{1/g}), (\protect\ref{exact}) and (\protect\ref{DW}), for large $g$ (and/or
small $m$), \textit{viz}., $g=5$, $m=0.152$.}
\label{fig:exact}
\end{figure}

In addition, Fig. \ref{fig:exact} displays the comparison of the analytical
approximation (TFA) based on Eqs. (\ref{1/g}), (\ref{exact}), and (\ref{DW})
with the numerical solution obtained for the same values of parameters. It
is observed that the analytically predicted density profiles are virtually
identical to their numerical counterparts.

\begin{figure}[th]
\centerline{(a) \hspace{8cm} (b)} \includegraphics[width=0.35%
\textwidth]{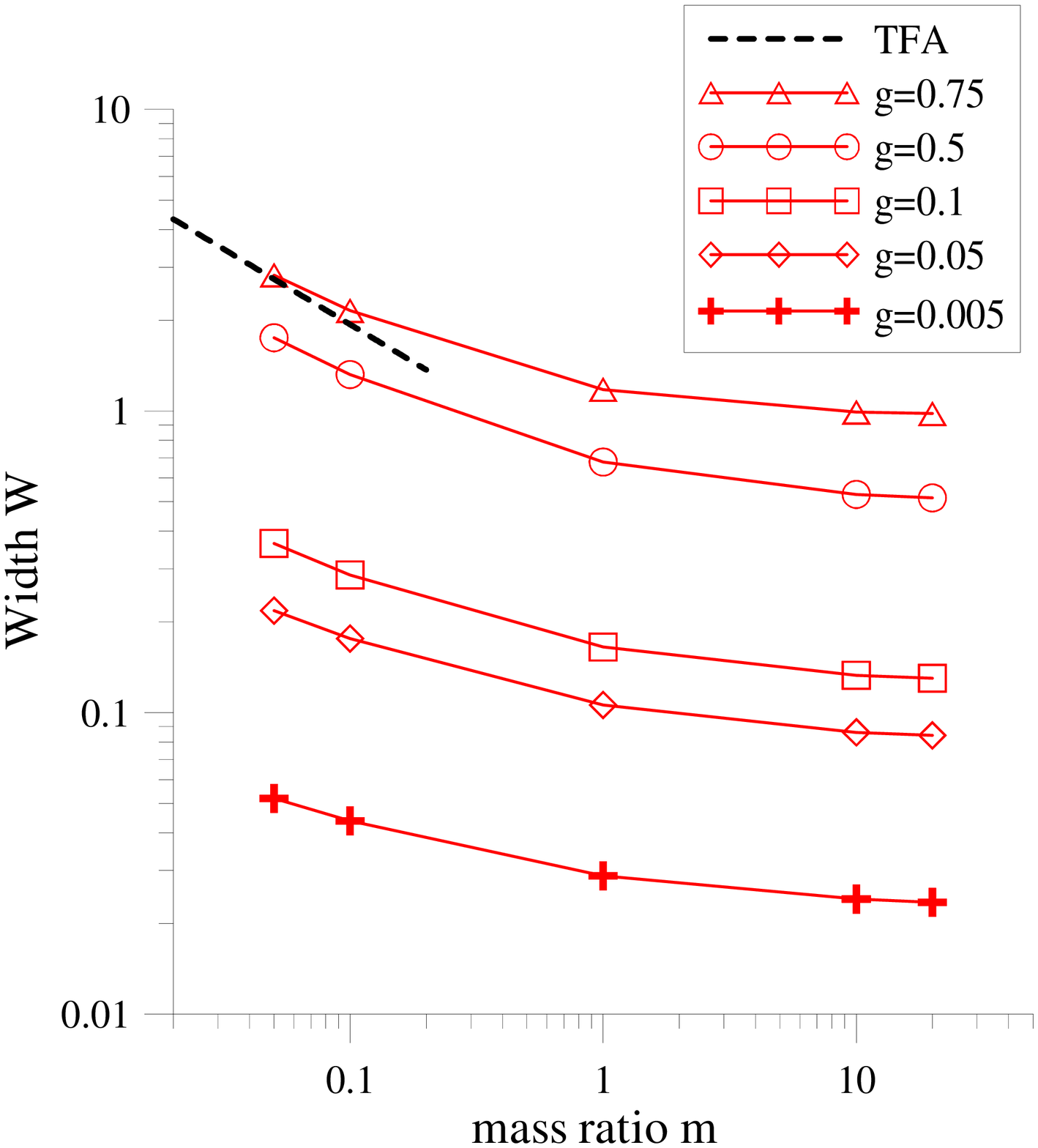} \hspace{2cm} \includegraphics[width=0.35%
\textwidth]{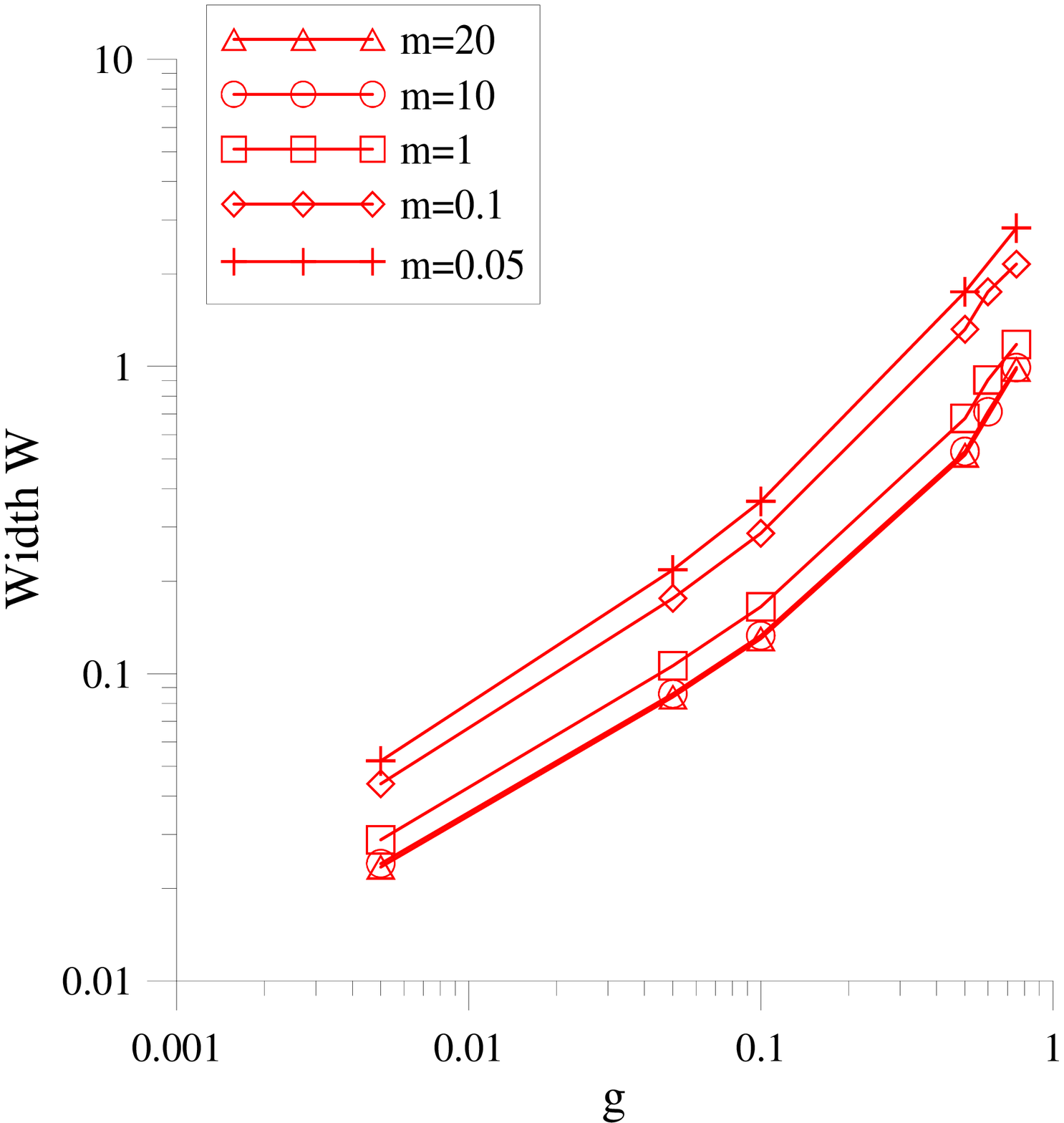}
\caption{(Color online) (a) The DW width, defined as per Eq. (\protect\ref{W}%
) vs. the TG/BEC (quintic/cubic) mass ratio, $m$, at different fixed values
of the interaction constant, $g$. The dashed black line shows the analytical
prediction (\protect\ref{WTFA})\ produced by the TFA for $g=3/4$. (b) The DW
width vs. $g$, at several fixed values of $m$. In both panels, $\protect\mu %
_{2}=1$ is fixed.}
\label{fig:Wvsm}
\end{figure}

To characterize the entire family of the DW solutions, their width defined
as per Eq. (\ref{W}) has been numerically evaluated, as shown in Fig. \ref%
{fig:Wvsm}(a), vs. the mass ratio, $m$, in the interval of $0.05\leq m\leq
50 $. It is seen that the dependence of the width on $m$ is rather weak, in
agreement with examples displayed in Fig. \ref{fig:DW}. Further, the
dependence of the DW's width on the BEC self-repulsion constant, $g$, is
shown in Fig. \ref{fig:Wvsm}(b). The range of values of $g$ displayed in the
figure is bounded by existence limits of the DW solutions, as given by Eq. (%
\ref{g<}). The increase of the width with $g$ is a natural property of the
system, as stronger self-repulsion stretches the transient layer [in the
analytical form, this is clearly shown by Eq. (\ref{broad}) and solution (%
\ref{DW}).]

\section{Bubble-drop (BD) states}

\subsection{Analytical considerations}

Bubble-drop (BD) solutions are even ones, with $\phi _{1,2}(-x)=\phi
_{1,2}(x)$, singled out by b.c.
\begin{gather}
\frac{d\phi _{1}}{dx}|_{x=0}=\frac{d\phi _{2}}{dx}|_{x=0}=0,  \notag \\
\phi _{1}\left( x=+\infty \right) =0,~\phi _{2}^{2}\left( x=+\infty \right)
\equiv \left( n_{2}\right) _{\mathrm{asympt}}=\sqrt{\mu _{2}}  \label{DB2}
\end{gather}%
(a BEC drop embedded into the TG background), or
\begin{eqnarray}
\frac{d\phi _{1}}{dx}|_{x=0} &=&\frac{d\phi _{2}}{dx}|_{x=0}=0,  \notag \\
\phi _{1}^{2}\left( x=+\infty \right) &\equiv &\left( n_{1}\right) _{\mathrm{%
asympt}}=\mu _{1}/g,~\phi _{2}\left( x=+\infty \right) =0  \label{DB1}
\end{eqnarray}%
(a TG drop embedded into the BEC background), cf. Eq. (\ref{bc}). In fact,
the BDs may be considered as bound pairs of the DWs, with the layer BEC or
TG trapped between the two semi-infinite TG or BEC domains, respectively.

The existence and stability of these solutions has been numerically checked
by fixing the chemical potential of the background to unity, and varying the
norm of the drop, or more precisely:

\begin{itemize}
\item setting $\mu _{2}=1$ for the TG background, see Eq. (\ref{DB2}), and
selecting several fixed values of $N_{1}$ for the BEC drop, see Eq. (\ref{N}%
);

\item setting $\mu _{1}=1$ for the BEC background, see Eq. (\ref{DB1}), and
selecting several fixed values of $N_{2}$ for the TG drop, see Eq. (\ref{N}).
\end{itemize}

It is relevant to stress that, unlike the DWs, the chemical potentials of
the bubble and drop components are not related by any condition similar to
Eq. (\ref{mumu}). In this setting, the remaining free parameters are
relative mass $m$ and BEC self-repulsion coefficient $g$.

Analytical results for the BD structures are available in the limit case of $%
m\rightarrow \infty $ (heavy TG atoms), which corresponds to Eqs. (\ref%
{phi2phi1}) and (\ref{phi1nophi2}). Indeed, in this approximation the BD
solutions, defined by b.c. (\ref{DB2}) or (\ref{DB1}) correspond,
respectively, to bright solitons or bubbles produced by Eq. (\ref{phi1nophi2}%
) (``bubbles" are solutions of NLSEs in the form of a local drop of the
density supported by the flat background, without zero crossing, and without
a phase shift at $x\rightarrow \pm \infty $, unlike dark solitons \cite%
{Barash}). Further, the analysis of the corresponding Hamiltonian (\ref%
{phi1h}) demonstrates that Eq. (\ref{phi1nophi2}) cannot generate bright
solitons, but it readily gives rise to bubbles, i.e., the BD patterns in the
form of the TG drop embedded into the BEC background. The fact that the BDs
exist in the form of the TG drop embedded into the BEC bubble, but do not
exist in the reverse form, is confirmed by numerical results presented below.

Analytical results for the BD structures are also available in the limit
case of $g\rightarrow \infty $ or $m\rightarrow 0$, when the derivative term
may be neglected in Eq. (\ref{phi1}), leading to Eq. (\ref{1/g}), as shown
above. The substitution of this approximation into Eq. (\ref{phi2}) leads to
the single equation with the cubic-quintic nonlinearity. For $\mu _{1}$ not
linked to $\mu _{2}$ by relation (\ref{mumu}), which was specific for the
DW, but is not relevant in the present case, the cubic-quintic equation
takes a form slightly more general than Eq. (\ref{single}) derived above:
\begin{equation}
\left( \mu _{2}-\frac{\mu _{1}}{g}\right) \phi _{2}=-\frac{1}{2m}\frac{%
d^{2}\phi _{2}}{dx^{2}}+\left( \phi _{2}^{4}-\frac{1}{g}\phi _{2}^{2}\right)
\phi _{2}.\;  \label{CQ}
\end{equation}%
A straightforward consideration demonstrates that Eq. (\ref{CQ}) gives rise
to bubbles, i.e., effectively, the BDs of type (\ref{DB2}), in the case of%
\begin{equation}
\frac{3}{8}<m\left( \frac{\mu _{1}}{g}-\mu _{2}\right) <\frac{1}{2},
\label{bubble}
\end{equation}%
and to bright solitons, i.e., the BDs of type (\ref{DB1}), in the range of%
\begin{equation}
0<m\left( \frac{\mu _{1}}{g}-\mu _{2}\right) <\frac{3}{8}.  \label{bright}
\end{equation}%
In fact, the present limit of $g\rightarrow \infty $ or $m\rightarrow 0$
implies that interval (\ref{bubble}) does not exist, while condition (\ref%
{bright}) may hold. Thus, this consideration again predicts that the BD
exist solely in the form of a TG\ drop (bright soliton) embedded into the
BEC bubble, which is exactly confirmed by numerical results following below.

In terms of the above-mentioned optics model, this conclusion means that a
bright soliton driven by the quintic self-defocusing may be embedded into
the background filled by the optical wave subject to the cubic
self-defocusing, but not vice versa.

\subsection{Numerical findings}

For the state defined as per Eq. (\ref{DB2}) (a BEC/cubic drop embedded into
the TG/quintic background), it was not possible to find solutions in the
entire parametric region investigated, which amounts to intervals $0.2\leq
g\leq 5$ and $0.1\leq m\leq 10$, and $0.1\leq N_{1}\leq 10$. In
imaginary-time simulations, this configuration, if taken as an input,
transforms into a flat one. The nonexistence of this BD species is readily
explained by the analytical results presented above.

\begin{figure}[tbph]
\includegraphics[width=0.35\textwidth]{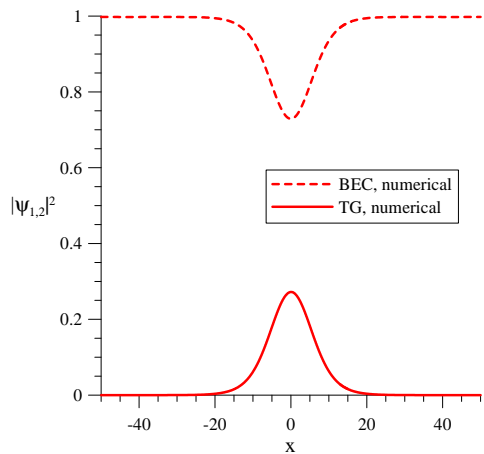}
\caption{(Color online) A typical example of a BD state, in the form of the
TG (quintic) drop with $N_{2}=4$, immersed into the BEC (cubic) background,
as produced by the numerical solution of Eqs. (\protect\ref{BEC}) and (%
\protect\ref{TG}) with b.c. given by Eq. (\protect\ref{DB1}). Other
parameters are $\protect\mu _{1}=1$, $g=1$, $m=1$, $\protect\mu _{2}=0.87$, $%
M_{\mathrm{void}} = 4$.}
\label{DB}
\end{figure}

\begin{figure}[th]
\includegraphics[width=0.35\textwidth]{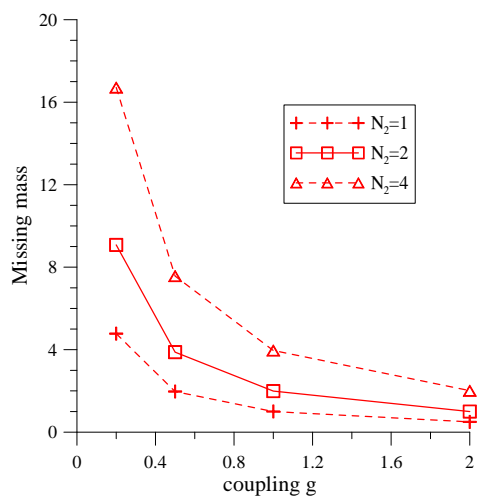}
\caption{(Color online) The missing mass in the bubble of the BEC gas (cubic
component) ousted by the TG (quintic) drop of norm $N_{2}$, as defined by
Eq. (\protect\ref{mvoid}), in the BD state. The BEC chemical potential is $%
\protect\mu _{1}=1$, and the TG/BEC mass ratio is $m=1$.}
\label{fig:missing_mass}
\end{figure}

On the other hand, also in agreement with the above analysis, in the same
parameter region (with $N_{1}$ replaced by $N_{2}$) stable solutions have
been readily found for the BD state defined by Eq. (\ref{DB1}) (a TG/quintic
drop embedded into the BEC/cubic background), see a typical example in Fig. %
\ref{DB}. The BD may be characterized by the \textit{missing mass} of the
bubble void, i.e.,
\begin{equation}
M_{\mathrm{void}}=\left[ \left( n_{1}\right) _{\mathrm{asympt}}\right]
^{-1}\int_{-\infty }^{+\infty }\left[ \left( n_{1}\right) _{\mathrm{asympt}%
}-\phi _{1}^{2}(x)\right] dx.  \label{mvoid}
\end{equation}%
Computations demonstrate that, for the family of the BD states, $M_{\mathrm{%
void}}$ very weakly depends on relative mass $m$. Figure \ref%
{fig:missing_mass} displays the missing mass of the BEC component as a
function of $g$. The dependence observed in Fig. \ref{fig:missing_mass} may
be approximated by relation $M_{\mathrm{void}}\simeq N_{2}/g$, which is easy
to understand: according to Eq. (\ref{phi1}), the TG component with density $%
\phi _{2}^{2}$ replaces the missing BEC field with effective density $\left(
\phi _{1}^{2}\right) _{\mathrm{void}}=\phi _{2}^{2}/g$. This argument also
explains an approximately linear increase of $M_{\mathrm{void}}$ with the
growth of the norm of the TG/quintic drop, $N_{2}$, as clearly seen in Fig. %
\ref{fig:missing_mass}.

\section{Dark-bright solitons (DBS)}

\subsection{The analytical formulation}

Bright-dark-soliton (DBS) solutions represent complexes with one even
localized (bright) component, and the other delocalized spatially odd
zero-crossing one (the dark soliton), the difference from the BD structure
being that in the latter case both components were patterned as even ones,
without zero crossing in the bubble structure. The corresponding b.c. for
the DBS complex are given by%
\begin{gather}
\frac{d\phi _{1}}{dx}|_{x=0}=\phi _{2}(x=0)=0,  \notag \\
\phi _{1}\left( x=+\infty \right) =0,~\phi _{2}^{2}\left( x=+\infty \right)
\equiv \left( n_{2}\right) _{\mathrm{asympt}}=\sqrt{\mu _{2}}  \label{DBS2}
\end{gather}%
(a bright BEC component embedded into the TG background), or
\begin{eqnarray}
\phi _{1}(x &=&0)=\frac{d\phi _{2}}{dx}|_{x=0}=0,  \notag \\
\phi _{1}^{2}\left( x=+\infty \right) &\equiv &\left( n_{1}\right) _{\mathrm{%
asympt}}=\mu _{1}/g,~n_{2}\left( x=+\infty \right) =0  \label{DBS1}
\end{eqnarray}%
(a bright TG component embedded into the BEC background), cf. Eqs. (\ref{DB2}%
) and (\ref{DB1}). Note that, similar to the case of the BD, the DBS
existence condition does not impose any relation on the chemical potentials
of its components, unlike Eq. (\ref{mumu}) for the DW.

The limit case of $m\rightarrow \infty $ (heavy TG atoms), which is
represented by Eqs. (\ref{phi2phi1}) and (\ref{phi1nophi2}), makes it
possible to produce analytical results for the DBS defined by b.c. (\ref%
{DBS1}), which corresponds to a dark soliton of Eq. (\ref{phi1nophi2}). The
analysis of the Hamiltonian (\ref{phi1h}) demonstrates that Eq. (\ref%
{phi1nophi2}) indeed gives rise to dark solitons, i.e., the DBS patterns in
the form of the bright TG soliton embedded into the dark BEC soliton.

The consideration of the DBS is also possible in the opposite limit of $%
g\rightarrow \infty $ or $m\rightarrow 0$, which amounts to the
consideration of Eq. (\ref{CQ}). Dark solitons of that equation correspond
to the DBS defined by b.c. (\ref{DBS2}), and it is easy to see that Eq. (\ref%
{CQ}) gives rise to the dark solitons in the range of $-\infty <m\left( \mu
_{1}/g-\mu _{2}\right) <3/8$. Obviously, this condition holds in the present
limit of  $g\rightarrow \infty $ or $m\rightarrow 0$.

Thus, the analysis of the limit cases readily predicts the existence of both
species of the DBSs, which correspond to b.c. (\ref{DBS2}) and (\ref{DBS1}).
This prediction is confirmed by the following numerical results.

\subsection{Numerical results}

\begin{figure}[th]
\centerline{(a) \hspace{6cm} (b)}
\includegraphics[width=0.35
\textwidth]{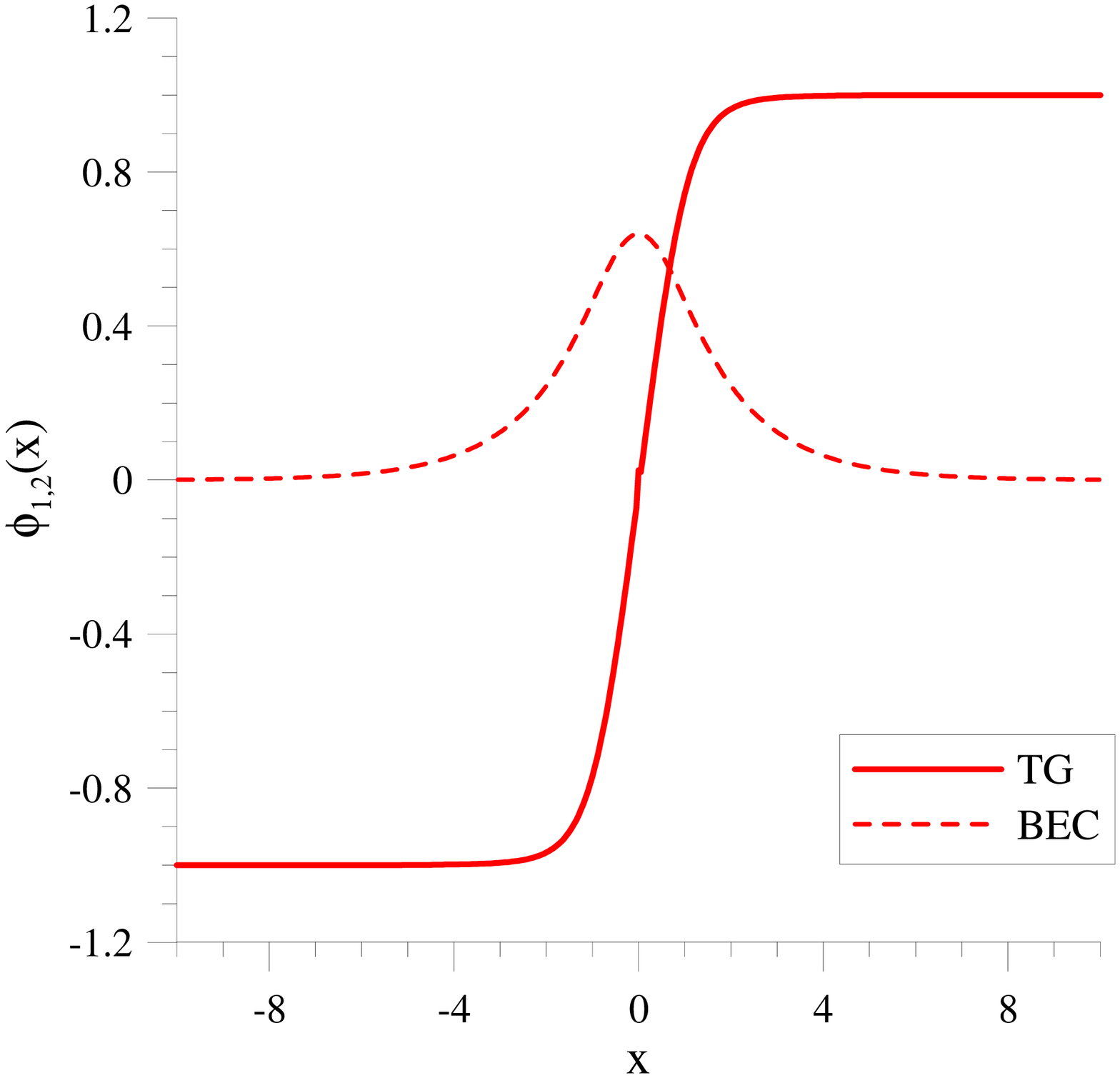} \includegraphics[width=0.35%
\textwidth]{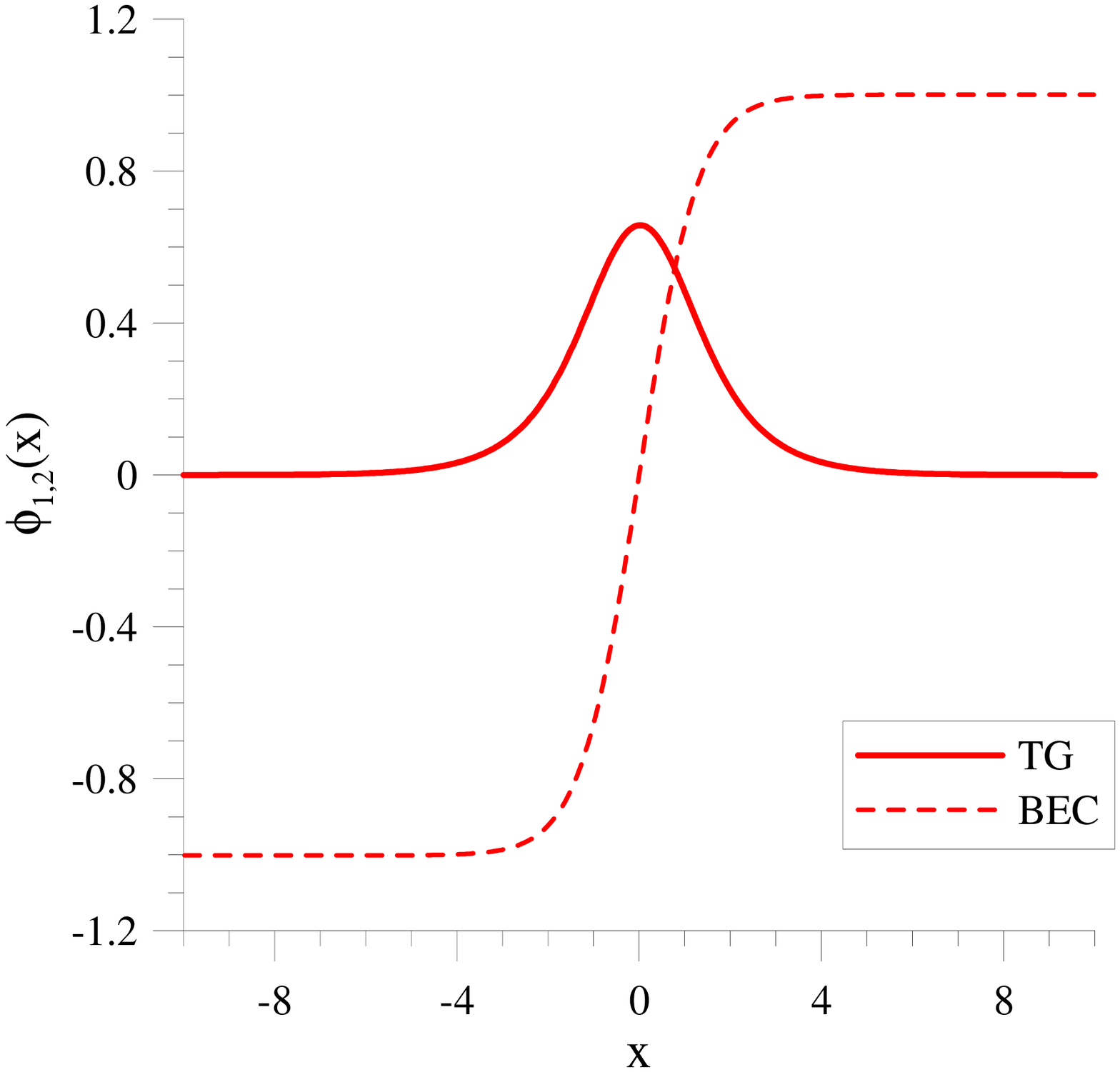}
\caption{(Color online) Generic profiles of BDS solutions. (a) The one with
the bright BEC (cubic) and dark TG (quintic) components, as per Eq. (\protect
\ref{DBS2}). (b) The one with the bright TG and dark BEC components, as per
Eq. (\protect\ref{DBS1}). Parameters are (a): $m=g=\protect\mu _{2}=1$, $%
N_{2}=1$, (b): $m=g=\protect\mu _{1}=1$, $N_{1}=1$.}
\label{fig:BDS_profile}
\end{figure}

Both types of the DBS have been produced by the numerical computations, as
shown in Fig. \ref{fig:BDS_profile}.

\begin{figure}[th]
\centerline{(a) \hspace{6cm} (b)} \includegraphics[width=0.35%
\textwidth]{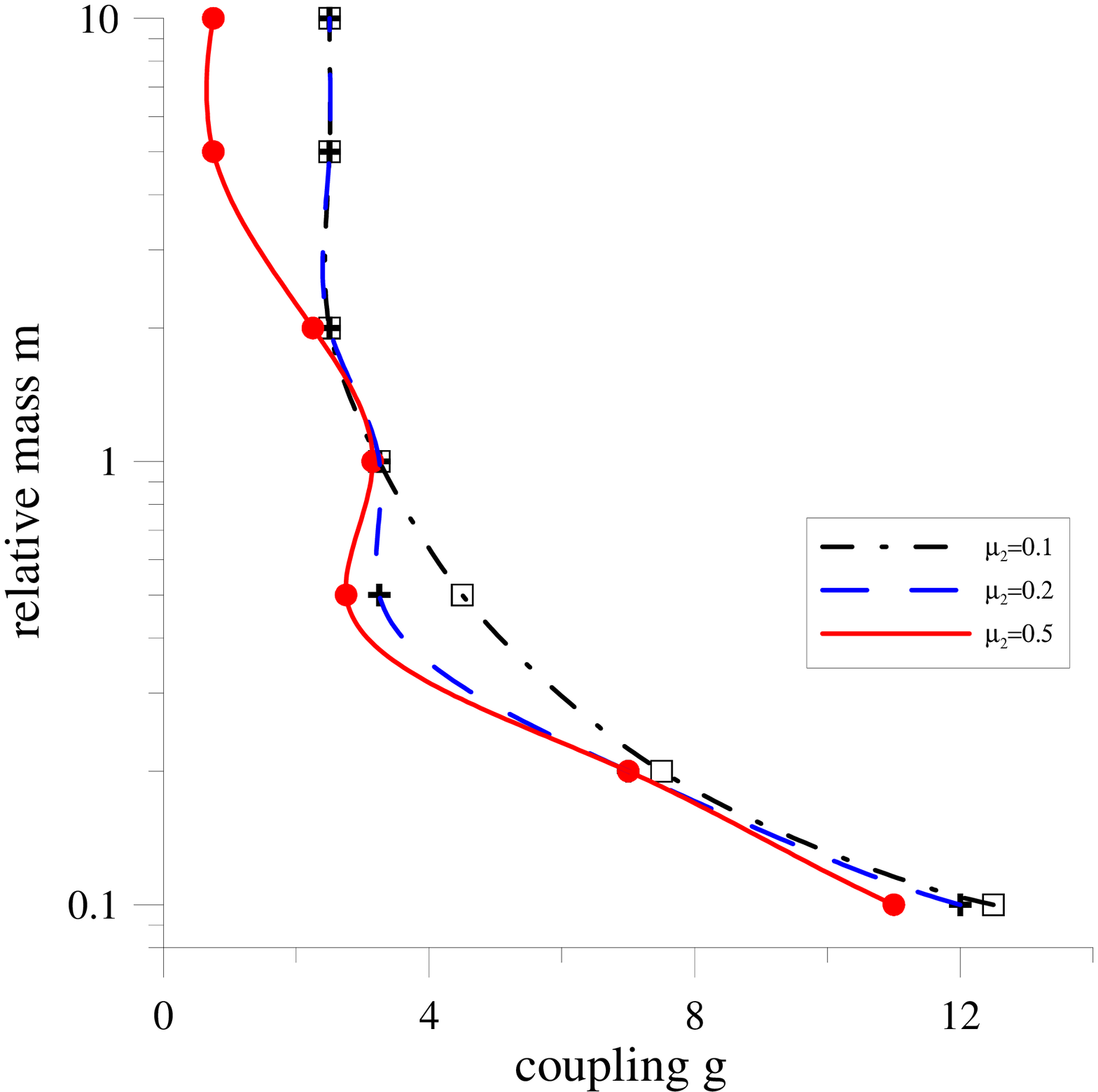}\hspace{.2cm} %
\includegraphics[width=0.35\textwidth]{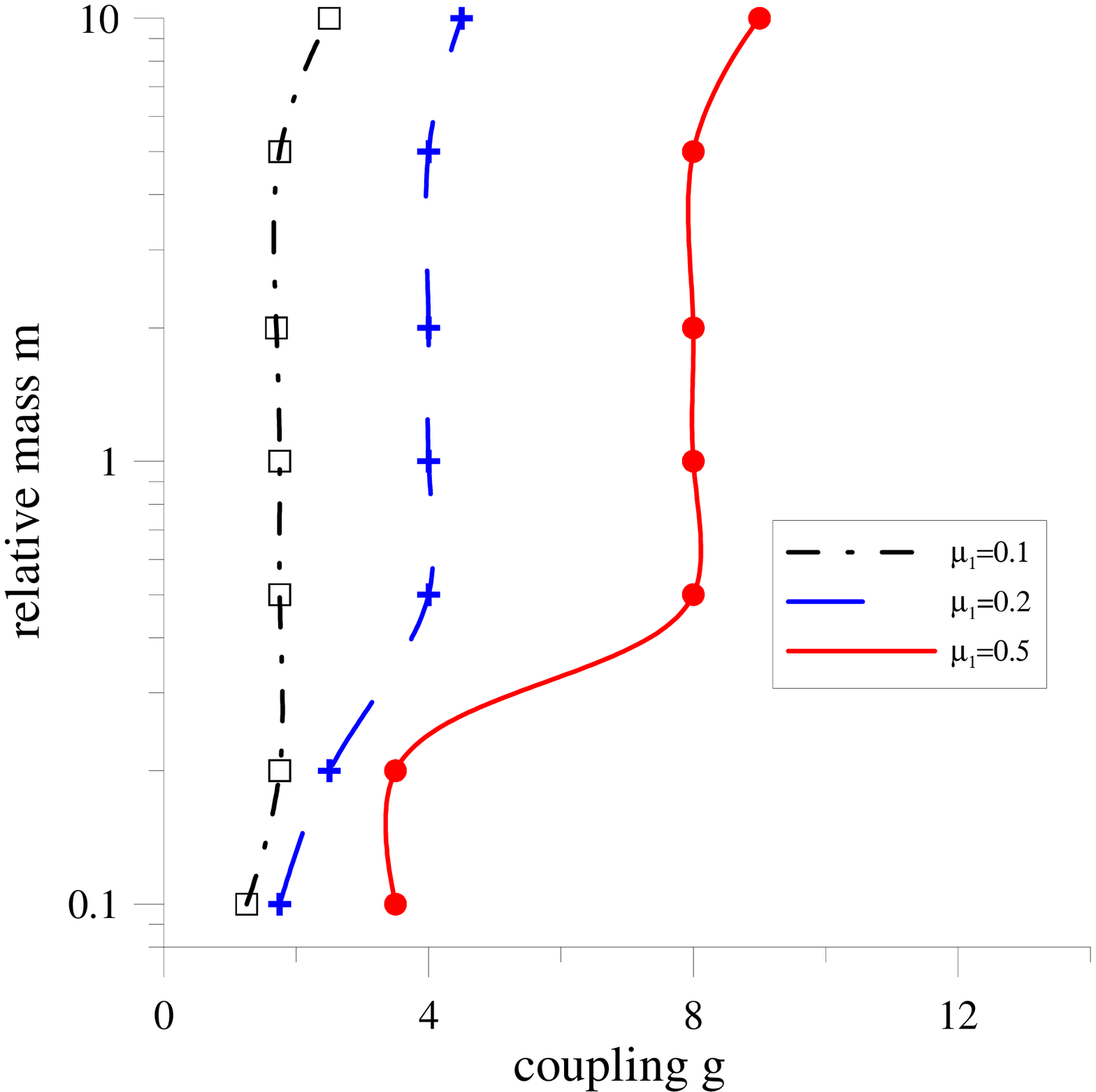}
\caption{(Color online) The BDSs, with the bright BEC (cubic) (a) orTG
(quintic) (b) component, exist on the left-hand side of the curves. Panels
(a) and (b) are drawn for $N_{1}=1$ and $N_{2}=1$, respectively.}
\label{fig:BDS_stability}
\end{figure}

The DBS complexes have their existence boundaries. They cease to exist when
the self-repulsion BEC coefficient, $g$, becomes too large. This is shown in
Fig. \ref{fig:BDS_stability}, that displays the boundaries in the plane of $%
\left( g,m\right) $. It is seen that a lower chemical potential of the dark
component favors the existence of the DBS, as the boundaries move towards
higher values of $g$ with the decrease of the chemical potential. Relative
mass $m$ plays a different role: the existence area for the DBS complex with
the bright BEC\ (cubic) component shrinks with the increase of $m$, while
for the case of the bright TG (quintic) component the increase of $m$ favors
the DBS existence.

\section{Conclusion}

We have introduced a binary system in the form of an immiscible pair of BEC
and TG quantum gases, in the effectively 1D setting. Using the system of the
GPE for the BEC, nonlinearly coupled to the quintic NLSE for the TG
component, we have analyzed the possibility of the existence of three types
of interfaces in this binary system: DWs (domain walls), BDs (bubble-drops),
and BDSs (bright-dark solitons). The same model applies to a bimodal light
propagation in a specially designed colloidal medium in optics. The
immiscibility condition for the binary system of the present type was
obtained in the general form. Analytical results were produced by means of
the TFAs (Thomas-Fermi approximations), in the combination with the
systematic numerical analysis. The DWs have been found an explicit
analytical form, by means of the TFA, and in the generic form numerically.
The analysis and numerical results demonstrate that the BDs exist solely in
the form of the TG drop immersed into the BEC background, while the BDS
complexes may be built equally well of the TG bright and BEC dark components
or vice versa. Thus, the predicted results suggest new experiments with
tightly confined two-species mixtures of ultracold bosonic gases. The
analysis can be extended for a chain of BDs and/or DBSs in a long system,
including a circular one, which corresponds to the binary gas loaded into a
tight toroidal trap \cite{ring-binary}.

\section*{Acknowledgements}

We appreciate valuable discussions with M. Salerno. G.F. acknowledges
financial support from Italian research programs PON Ricerca e Competitivit%
\`{a} 2007-2013, under grant agreement PON NAFASSY, PONa3\_00007, and
Programma regionale per lo sviluppo innovativo delle filiere Manifatturiere
strategiche della Campania - Filiera WISCH Progetto 2: Ricerca di tecnologie
innovative digitali per lo sviluppo sistemistico di computer, circuiti
elettronici e piattaforme inerziali ad elevate prestazioni ad uso avionico.

\end{document}